\documentclass[]{spie}  

\usepackage{amsmath,amsfonts,amssymb}
\usepackage{graphicx}
\usepackage[colorlinks=true, allcolors=blue]{hyperref}
\usepackage{subcaption}
\usepackage{comment}
\usepackage[dvipsnames]{xcolor}

\title{The Blue Multi Unit Spectroscopic Explorer (BlueMUSE) on the VLT: characterization of two VPHG prototypes based on dichromated gelatin and photopolymer recording materials}

\author[a,*]{Alexandre Jeanneau}
\author[b]{Andrea Bianco}
\author[c]{Andrew Clawson}
\author[b]{Michele Frangiamore}
\author[c]{Elroy Pearson}
\author[d]{Laurent Pinard}
\author[e]{Jürgen Schmoll}
\author[a]{Johan Richard}
\author[a]{Rémi Giroud}
\author[a]{Florence Laurent}
\author[a]{Roland Bacon}

\affil[a]{Université Lyon, Université Claude Bernard Lyon 1, Ens de Lyon, CNRS, Centre de Recherche Astrophysique de Lyon, F-69230, Saint-Genis-Laval, France}
\affil[b]{INAF – Osservatorio Astronomico di Brera, via E. Bianchi 46, 23807, Merate, LC, Italy}
\affil[c]{Wasatch Photonics, 1305 North 1000 West, Suite 120, Logan, UT 84321 USA}
\affil[d]{Université Lyon, Université Claude Bernard Lyon 1, CNRS, Laboratoire des Matériaux Avancés, F-69622, Villeurbanne, France}
\affil[e]{Centre for Advanced Instrumentation (CfAI), Durham University, South Road, Durham DH1 3LE, United Kingdom}

\authorinfo{Send correspondence to alexandre.jeanneau@univ-lyon1.fr}

\begin{document} 
\maketitle

\begin{abstract}
Volume-phase holographic gratings (VPHGs) are widely used in astronomical spectrographs due to their adaptability and high diffraction efficiency.
Most VPHGs in operation use dichromated gelatin as a recording material, whose performance is sensitive to the coating and development process, especially in the near-UV.
In this letter, we present the characterization of two UV-blue VPHG prototypes for the BlueMUSE integral field spectrograph on the VLT, based on dichromated gelatin and the Bayfol®HX photopolymer film as recording materials.
Our measurements show that both prototypes meet the required diffraction efficiency and exhibit similar performance with a wavelength-average exceeding 70\% in the 350-580 nm range. Deviations from theoretical models increase towards 350 nm, consistently with previous studies on similar gratings. We also report similar performances in terms spatial uniformity and grating-to-grating consistency. Likewise, no significant differences in wavefront error or scattered light are observed between the prototypes. 
\end{abstract}


\section{Introduction}
\label{sect:intro}
Volume-phase holographic gratings (VPHGs) operate through bulk refractive index modulations, which are created by holographically exposing a photosensitive material such as dichromated gelatin (DCG). Despite their widespread use in astronomical spectrographs over the past decades — owing to their adaptability and high diffraction efficiency\cite{Barden98, Barden00} — the choice of VPHG suppliers is rather limited. This is partly due to the sensitivity of the DCG coating and development process\cite{Blanche04}, particularly towards the near-UV. Indeed, measurements of 36 VPHGs for the Dark Energy Spectroscopic Instrument (DESI) suggest higher processing-induced efficiency variations in the UV–blue gratings with respect to other bands\cite{Ishikawa18}. Measurements of four VPHG prototypes for the Visible Integral field Replicable Unit Spectrograph (VIRUS) reveal increasing efficiency deviations to the theoretical model towards 350 nm, which are attributed to process-induced absorption or scatter in the DCG layer\cite{Chonis12}. However, Ref.~\citenum{Chonis14} reports both minimal grating-to-grating scatter and minimal spatial variations towards 350 nm for the full suite of 170 VIRUS VPHGs.

Laminated self-processing photopolymers seem well-suited to facilitate VPHG manufacturing, and offer an increasingly broad range of thicknesses and achievable refractive index modulations\cite{Bianco23}. This solution is particularly interesting for near-UV/visible instruments such as BlueMUSE\cite{Richard24} on the VLT, which requires a high and consistent transmission across 16 replicated integral-field units.

In this letter we compare two VPHG prototypes for BlueMUSE based on dichromated gelatin (DCG) and the Bayfol®HX photopolymer film\cite{Bruder17} as recording materials. We assess manufacturing losses between modeled and measured diffraction efficiencies and cross-check characterization results provided by the manufacturers using an identical full aperture test setup. We also compare the transmitted wavefront error as well as the bidirectional transmittance distribution function.

\section{Method}
\label{sect:method}

\subsection{Manufacturing}
\begin{figure}[ht!] 
\centering
\includegraphics[height=4cm]{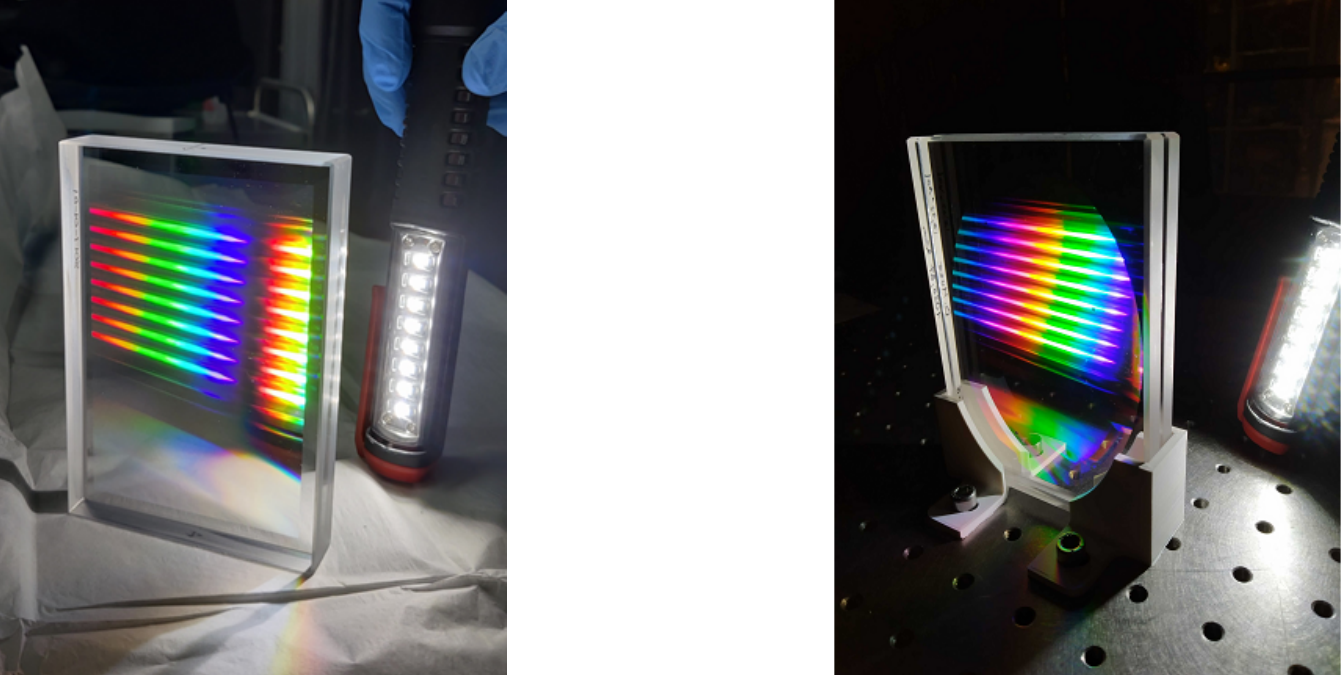}
\\
(a) \hspace{4.2cm} (b)
\caption{DCG (a) and photopolymer (b) prototypes, back illuminated by a handheld LED light.}
\label{fig:photos} 
\end{figure}

Two VPHG prototypes are manufactured according to the design parameters listed in Table~\ref{tab:parameters}: the DCG-based prototype is manufactured by Wasatch Photonics and another prototype based on the Bayfol®HX photopolymer is manufactured by the Osservatorio Astronomico di Brera (OAB). The resulting gratings are shown in Figure~\ref{fig:photos}.

\begin{table*}[ht!]
\centering
\caption{Design parameters.}
\label{tab:parameters}
\begin{tabular}{lr}
\hline
\hline
\noalign{\smallskip}
Design parameter & Value \\
\noalign{\smallskip}
\hline
\noalign{\smallskip}
Wavelength range &  350 - 580 nm \\
\noalign{\smallskip}
Angle of incidence (AOI, in air) &  $13.72$° $\pm$ $0.7$° \\
 \noalign{\smallskip}
Line density &  $1027 \pm 1$ line/mm\\
 \noalign{\smallskip}
Clear aperture &  $110 \times 70$ mm ellipse \\
 \noalign{\smallskip}
Diffraction efficiency &  $T(350\text{ nm}) \gtrsim 60\%$\\
(excl. Fresnel losses) &  $T(580\text{ nm}) > 40\%$\\
 &  Goal: $T_\text{avg} > 70\%$\\
 \noalign{\smallskip}
Wavefront error (+1 order) &  $< 1266$ nm ($2\lambda$) PV\\
 \noalign{\smallskip}
Substrate &  Fused Silica (uncoated)\\
 \noalign{\smallskip}
Substrate size &  $130 \times 90$ mm \\
\noalign{\smallskip}
\hline
\end{tabular}
\end{table*}

\subsection{Diffraction efficiency test}
\begin{figure}
\centering
\includegraphics[height=4cm]{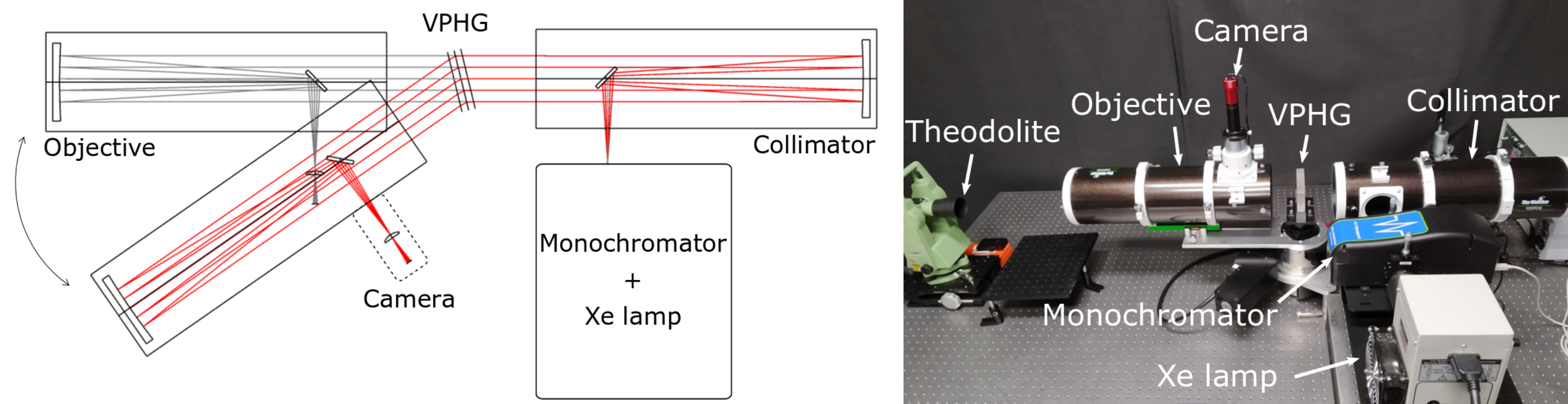}
\vspace{2mm}
\caption{Diffraction efficiency setup at CRAL.}
\label{fig:DE_setup} 
\end{figure}

Independent diffraction efficiency measurements have been carried out by Wasatch Photonics, OAB and the Centre de Recherche Astrophysique de Lyon (CRAL). Unlike the setups at Wasatch Photonics and OAB which rely on scanning a monochromatic test beam over the aperture and across the wavelength range, the method at CRAL is based on a full aperture test setup shown in Figure~\ref{fig:DE_setup} and akin to the one described in Ref~\citenum{Barkhouser14}. This choice was motivated by the desire to grasp diffraction efficiency variations over the clear aperture.

Briefly, the test beam is generated by a 150 W Xenon light source (Newport \#6256 arc lamp) and filtered by a monochromator (Oriel  Cornerstone CS130B) to a bandwidth of 4.5 nm varied from 330 to 600 nm. The beam is then directed towards a $f=650$ mm -- F/5 Newtonian telescope (Skywatcher Explorer 130PDS), used as a collimator. We mount the grating on a rotary stage and set it perpendicular to the collimated beam, using an autocollimation off the uncoated substrate. We establish this reference angle using a theodolite aligned with the monochromator slit. The grating is then rotated to the desired angle of incidence and an identical telescope, used as an objective, is rotated as well to collect the diffracted light. A CCD camera (ZWO ASI174MM Mini) is placed after the telescope focus in order to image the grating clear aperture.

We divide the diffracted image by a reference image, where the grating is removed and both telescopes are aligned. In order to increase the overall S/N, we stack 5 exposures per diffracted and reference image. Finally, we smooth the resulting diffraction efficiency map using a $1 \times 1$ mm boxcar filter. We assess the accuracy of this setup to $\sim 2\%$ using repeated transmission measurements of an uncoated fused silica window. The setup is probably limited by straylight and lamp stability between the diffracted and reference images, which are separated by less than a minute.

\section{Results}
\label{sect:results}

\subsection{Diffraction efficiency}
\begin{figure}
\centering
\includegraphics[width=\linewidth]{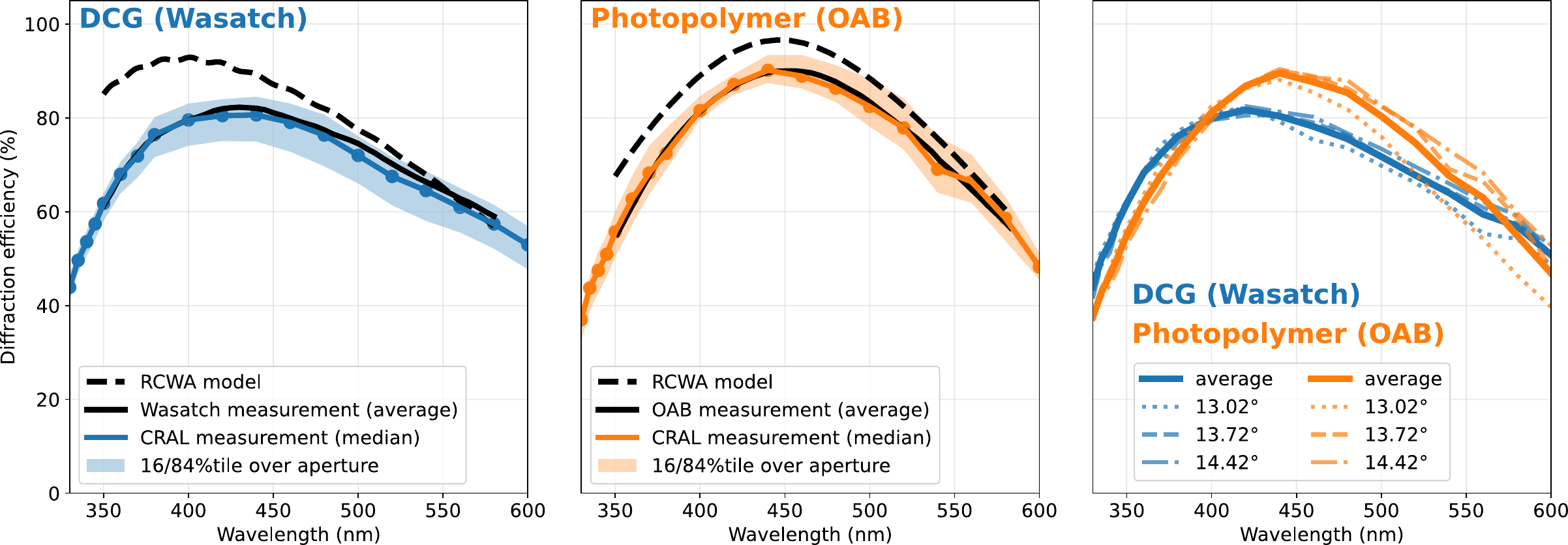}
\\
(a) \hspace{4.7cm} (b) \hspace{4.7cm} (c)
\caption{Diffraction efficiency results: spatial uniformity of the DCG (a) and photopolymer (b) prototypes (AOI = 13.72°), and angular selectivity (c).}
\label{fig:DE} 
\end{figure}

Internal diffraction efficiency results (i.e., corrected for Fresnel losses) for unpolarized light are presented in Figure~\ref{fig:DE}. Both prototypes comply with the requirements presented in Section \ref{sect:method}, albeit with different optimizations: the DCG prototype exhibits a lower average (72\% versus 76\%) but a better diffraction efficiency at 350 nm (62\% versus 55\%). We emphasize that average diffraction efficiency and diffraction efficiency at 350 nm were given the same importance in the optimization of those prototypes, considering the combined impact of atmosphere cutoff and internal glass absorption on the overall transmission of BlueMUSE. We observe that measurements at CRAL agree with data from Wasatch Photonics and OAB to within 1\% rms.

We note that both prototypes almost reach the performance predicted by Rigorous Coupled Wave Analysis (RCWA) models\cite{Moharam81} at the red end but increasingly deviate towards the blue, particularly for the DCG prototype. The overall deviation for an angle of incidence of 13.72° is 23\% for DCG and 12\% for Bayfol®HX at 350 nm, compared to 9\% and 7\% on average, respectively. Finally, we find 16\textsuperscript{th}-84\textsuperscript{th} percentile variations ($\pm 1\sigma$) of 6\% across the clear aperture for DCG and 10\% for Bayfol®HX at 350 nm, compared to 9\% on average over the full wavelength range for both prototypes.

\subsection{Wavefront error}
\begin{figure}
\centering
\includegraphics[width=\linewidth]{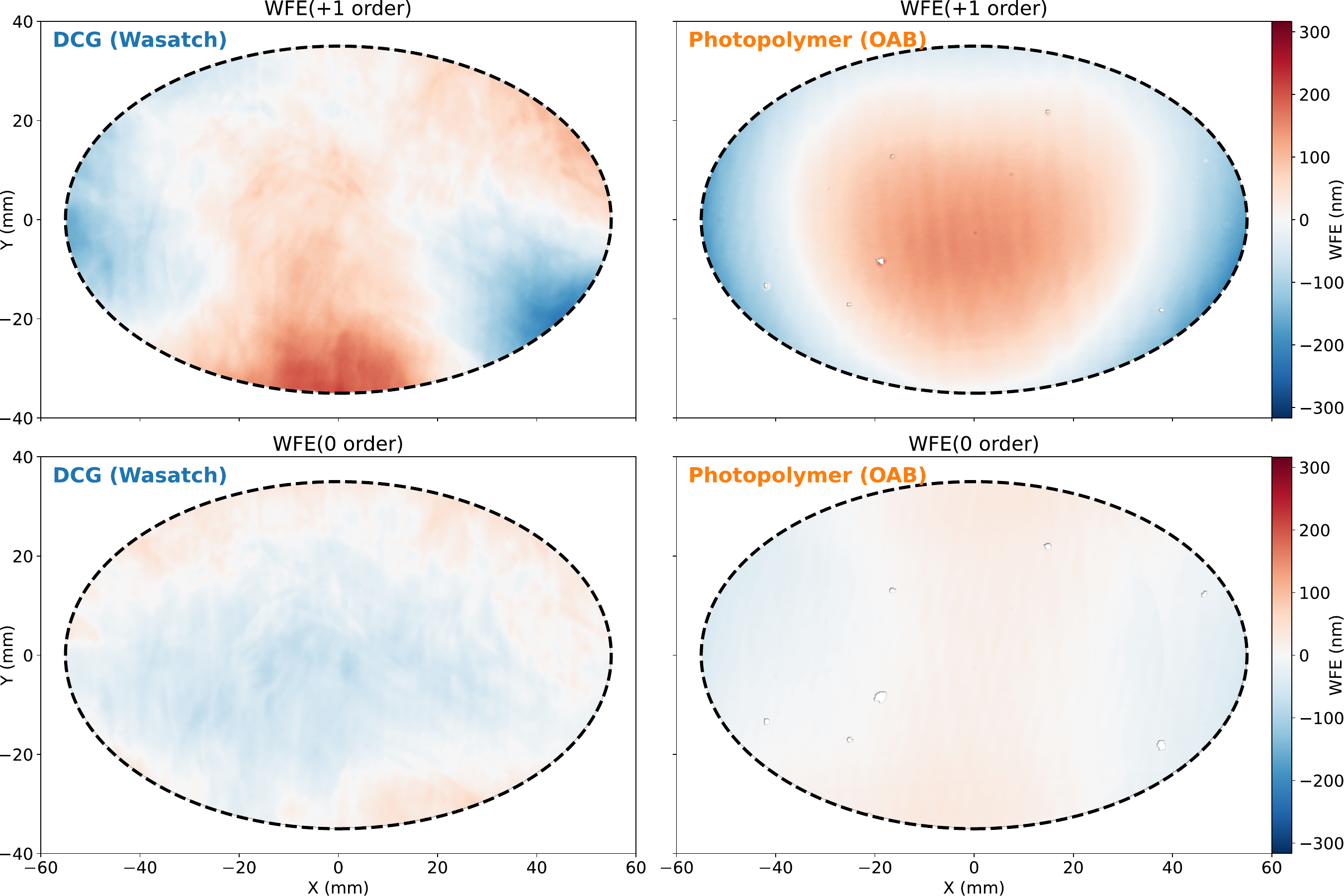}
\caption{Wavefront error (WFE) maps.}
\label{fig:WFE} 
\end{figure}

We measure the transmitted wavefront error in 0\textsuperscript{th} and +1 order using a Fizeau interferometer (4" Zygo Verifire XPD, $\lambda = 632.8$ nm) in a double-pass configuration, with the prototype set at its working angle and a $\lambda / 25$ flat mirror closing the cavity. As the prototype clear aperture is slightly larger than the 4-inch test beam, we stitch three measurements per diffraction order. The resulting wavefront error maps are shown in Figure \ref{fig:WFE}. Both the DCG and photopolymer prototypes are well within the required wavefront error ($< 2\lambda$ PV) with 451 nm PV (70 nm rms) and 523 nm PV (78 nm rms), respectively.

\subsection{Scattering}
\begin{figure}[ht!] 
\centering
\includegraphics[width=0.95\linewidth]{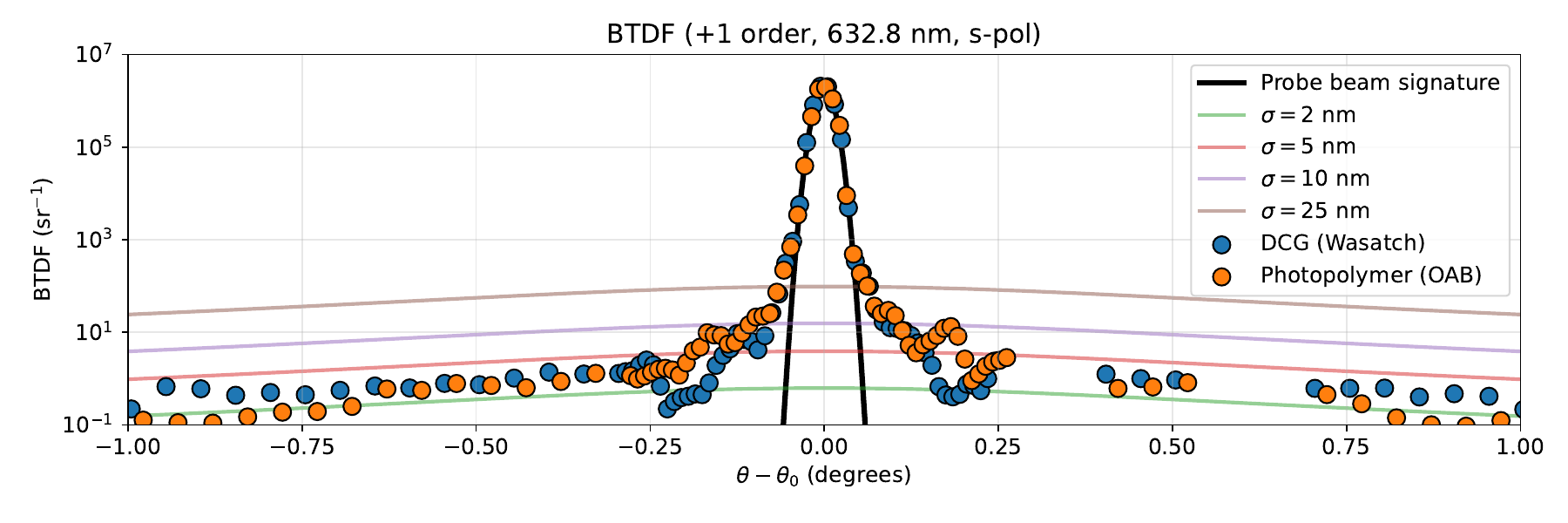}
\caption{Measured BTDFs in the dispersion direction, compared with the probe beam signature (black line) and models of smooth air-glass interfaces with varying microroughness (colored lines). Internal reflections at $\sim$0.3° and $\sim$0.6° are not shown.}
\label{fig:BTDF}
\end{figure}

Finally, we measure the Bidirectional Transmittance Distribution Function (BTDF) in the dispersion direction and at the center of the clear aperture using a complete angle scatter instrument (TSW CASI scatterometer). The probe beam is a s-polarized HeNe laser with a $\sim$ 1 mm spot diameter. The measured BTDF is shown in Figure~\ref{fig:BTDF}. Both prototypes have similar BTDFs within the limits of the probe beam signature, which spans $\sim$10 BlueMUSE pixels compared to a slit width of 2 pixels. Therefore, this measurement mainly probes the outer scatter halo surrounding the point spread function. 

\section{Discussion}
\label{sect:discussion}

\subsection{Diffraction efficiency uniformity and consistency}
Our measurements are consistent with expectations from Ref.~\citenum{Chonis12} which reports increasing deviations from RCWA models towards the blue, with deviations up to 10-30\% ($\sim$20\% median) at 350 nm, for DCG-based gratings with a similar bandpass. According to Ref.~\citenum{Chonis12}, these deviations are likely due to processing-related absorption or scatter, as they vary between production batches. The photopolymer prototype is at the lower end, with a 12\% deviation at 350 nm, compared to a median value of 23\% for the DCG prototype.

Although photopolymers are anticipated to provide more uniform performance across the clear aperture, our measurements show similar spatial variations for both the photopolymer and DCG prototypes. Notably, spatial variations above the median are smaller than those below the median, particularly for the DCG prototype. This aligns with Ref.~\citenum{Chonis14}, which attributes this to the higher likelihood of encountering a less favorable set of recording material properties (e.g., thickness or refractive index modulation) compared to the set optimized by design. Interestingly, the photopolymer prototype shows more balanced spatial variations around the median, even at peak diffraction efficiency (i.e., close to an optimal set of recording material properties). However, it is difficult to draw broad conclusions from a single realization of each manufacturing process.

To investigate grating-to-grating uniformity, OAB conducted a process repeatability test, recording and measuring four times a full aperture photopolymer film before encapsulation. The achieved diffraction efficiency demonstrated remarkable consistency, with wavelength-averaged variations of only 2\%. For comparison, similar studies such as Ref.~\citenum{Ishikawa18} and Ref.~\citenum{Chonis14} report grating-to-grating variations of 7\% rms and 3-6\% rms respectively, for comparable but fully assembled DCG-based gratings. This consistency suggests that most of the spatial variations arise from very repeatable edge effects. Indeed, we observe that most of the efficiency variation is located at the edge of the clear aperture as in Ref.~\citenum{Ishikawa18}.

\subsection{Wavefront error}
We note that the wavefront error in 0\textsuperscript{th} order is 3-4 times smaller than in the +1 order. This suggests that hologram recording errors are the primary contributors to the overall wavefront error, outweighing errors due to the substrate, recording material, or bonding, which tend to affect both orders similarly. In the photopolymer prototype, a significant portion of the diffracted wavefront error is due to a power term, which may be corrected during spectrograph alignment. This error likely arises from a slight collimation error in the recording beam.

\subsection{Scattering}
We approximate the measured straylight levels presented above by an equivalent surface microroughness, a useful quantity for generating simple yet realistic straylight models. We compare the prototypes against model BTDFs for smooth air-glass interfaces ($\Delta n = 0.5$) of varying microroughness ($\sigma = 2, 5, 10, 25$ nm), following the methodology of Ref~\citenum{Harnisch17}. We find an equivalent microroughness between 25 and 10 nm up to 0.10°, between 10 and 5 nm up to 0.25°, and below 5 nm for angles larger than 0.25°. This is a good straylight performance compared to the different grating technologies with similar line densities tested in Ref~\citenum{Harnisch17}, including a DCG-based VPHG which reaches an equivalent microroughness below 5 nm for angles larger than 1.5°.

\section{Summary and conclusions}
\label{sect:conclusions}
In this letter we have compared two UV-blue VPHG prototypes developped for BlueMUSE, based on dichromated gelatin (DCG) and the Bayfol®HX photopolymer as recording materials:
\begin{enumerate}
\item We have presented a full aperture test setup at CRAL, which provides measurements in agreement to within 1\% rms with respect to data from Wasatch Photonics and OAB.
\item Both prototypes comply with the required diffraction efficiency and achieve similar performance exceeding the wavelength-averaged goal of 70\% over the BlueMUSE bandpass.
\item We observe that both prototypes increasingly depart from their RCWA model towards 350 nm, which confirms a trend reported in Ref.~\citenum{Chonis12} for VIRUS prototypes.
\item Furthermore, we note that both prototypes have $\sim 10$\% spatial efficiency variations. A repeatability test at OAB shows a remarkably consistent grating-to-grating performance with wavelength-averaged deviations of only 2\%.
\item No significant differences in terms of wavefront error ($<1\lambda$ PV) or scattered light are observed.
\end{enumerate}

Although our measurements are based on a single realization for each manufacturing process, they offer valuable insights into the expected performance of the full suite of 16 VPHGs for BlueMUSE. Future statistical analyses of multiple photopolymer-based VPHGs will provide a more comprehensive understanding of the advantages and disadvantages of this recording material compared to DCG.

\acknowledgments 

We acknowledge financial support from the Commission spécialisée Astronomie Astrophysique (CSAA) of CNRS / INSU. We also acknowledge support from the FRAMA (Fédération de Recherche André-Marie Ampère) as well as the Labex-LIO (Lyon Institute of Origins) under Grant No. ANR-10-LABX-66 (Agence Nationale pour la Recherche).

\bibliography{report} 

\begin{thebibliography}{10}

\bibitem{Barden98}
Barden, S.~C., Arns, J.~A., and Colburn, W.~S., ``{Volume-phase holographic
  gratings and their potential for astronomical applications},'' in [{\em
  Optical Astronomical Instrumentation}{\nolinebreak\hspace{0.1em}]},
  D'Odorico, S., ed.,  {\bf 3355},  866 -- 876, International Society for
  Optics and Photonics, SPIE (1998).

\bibitem{Barden00}
Barden, S.~C., Arns, J.~A., Colburn, W.~S., and Williams, J.~B.,
  ``Volume‐phase holographic gratings and the efficiency of three simple
  volume‐phase holographic gratings,'' {\em Publications of the Astronomical
  Society of the Pacific}~{\bf 112},  809 (jun 2000).

\bibitem{Blanche04}
Blanche, P.-A., Gailly, P., Habraken, S.~L., Lemaire, P.~C., and Jamar, C.
  A.~J., ``{Volume phase holographic gratings: large size and high diffraction
  efficiency},'' {\em Optical Engineering}~{\bf 43}(11),  2603 -- 2612 (2004).

\bibitem{Ishikawa18}
Ishikawa, Y., Sirk, M.~M., Edelstein, J., Jelinsky, P., Brooks, D., Tarle, G.,
  and Collaboration), D., ``Comprehensive measurements of the volume-phase
  holographic gratings for the {D}ark {E}nergy {S}pectroscopic {I}nstrument,''
  {\em The Astrophysical Journal}~{\bf 869},  24 (dec 2018).

\bibitem{Chonis12}
Chonis, T.~S., Hill, G.~J., Clemens, J.~C., Dunlap, B., and Lee, H., ``{Methods
  for evaluating the performance of volume phase holographic gratings for the
  VIRUS spectrograph array},'' in [{\em Ground-based and Airborne
  Instrumentation for Astronomy IV}{\nolinebreak\hspace{0.1em}]},  McLean,
  I.~S., Ramsay, S.~K., and Takami, H., eds.,  {\bf 8446},  84465H,
  International Society for Optics and Photonics, SPIE (2012).

\bibitem{Chonis14}
Chonis, T.~S., Frantz, A., Hill, G.~J., Clemens, J.~C., Lee, H., Tuttle, S.~E.,
  Adams, J.~J., Marshall, J.~L., DePoy, D.~L., and Prochaska, T., ``{Mass
  production of volume phase holographic gratings for the VIRUS spectrograph
  array},'' in [{\em Advances in Optical and Mechanical Technologies for
  Telescopes and Instrumentation}{\nolinebreak\hspace{0.1em}]},  Navarro, R.,
  Cunningham, C.~R., and Barto, A.~A., eds.,  {\bf 9151},  91511J,
  International Society for Optics and Photonics, SPIE (2014).

\bibitem{Bianco23}
Bianco, A., Frangiamore, M., Zanutta, A., Oggioni, L., Pariani, G., Garzon, F.,
  and Insausti, M., ``{Improvements in VPHGs for astronomy based on
  photopolymers},'' in [{\em Holography: Advances and Modern Trends
  VIII}{\nolinebreak\hspace{0.1em}]},  Fimia, A. and Hrabovsk{\'y}, M., eds.,
  {\bf 12574},  125740H, International Society for Optics and Photonics, SPIE
  (2023).

\bibitem{Richard24}
Richard, J., Giroud, R., Laurent, F., Krajnovi\'c, D., Jeanneau, A., Bacon, R.,
  Abreu, M., Adamo, A., Araujo, R., Bouché, N., Brinchmann, J., Cai, Z.,
  Castro, N., Calcines, A., Chapuis, D., Claeyssens, A., Cortese, L., Daddi,
  E., Davison, C., Harris, R., Hayes, M., Jauzac, M., Kelz, A., Kneib, J.-P.,
  Goodwin, M., Lanotte, A.~A., Lawrence, J., Bouteiller, V.~L., Breton, R.~L.,
  Lehnert, M., Sanchez, A.~L., McGregor, H., McLeod, A.~F., Monteiro, M.,
  Morris, S., Opitom, C., Pécontal, A., Robertson, D., Smith, R., Steinmetz,
  M., Swinbank, M., Urrutia, T., der Sande, J.~V., Verhamme, A., Weilbacher,
  P.~M., Wendt, M., Wildi, F., and Zheng, J., ``The {B}lue {M}ulti {U}nit
  {S}pectroscopic {E}xplorer ({BlueMUSE}) on the {VLT}: science drivers and
  overview of instrument design,'' in [{\em Ground-based and Airborne
  Instrumentation for Astronomy X}{\nolinebreak\hspace{0.1em}]},  Bryant,
  J.~J., Motohara, K., and Vernet, J. R.~R., eds., {\em Proc. SPIE} {\bf 13096}
  (2024).

\bibitem{Bruder17}
Bruder, F.-K., Fäcke, T., and Rölle, T., ``The chemistry and physics of
  {Bayfol® HX} film holographic photopolymer,'' {\em Polymers}~{\bf 9}(10)
  (2017).

\bibitem{Barkhouser14}
Barkhouser, R.~H., Arns, J., and Gunn, J.~E., ``{Volume phase holographic
  gratings for the Subaru Prime Focus Spectrograph: performance measurements of
  the prototype grating set},'' in [{\em Ground-based and Airborne
  Instrumentation for Astronomy V}{\nolinebreak\hspace{0.1em}]},  Ramsay,
  S.~K., McLean, I.~S., and Takami, H., eds.,  {\bf 9147},  91475X,
  International Society for Optics and Photonics, SPIE (2014).

\bibitem{Moharam81}
Moharam, M.~G. and Gaylord, T.~K., ``Rigorous coupled-wave analysis of
  planar-grating diffraction,'' {\em J. Opt. Soc. Am.}~{\bf 71},  811--818 (Jul
  1981).

\bibitem{Harnisch17}
Harnisch, B., Deep, A., Vink, R., and Coatantiec, C., ``{Grating scattering
  BRDF and imaging performances: A test survey performed in the frame of the
  flex mission},'' in [{\em International Conference on Space Optics — ICSO
  2012}{\nolinebreak\hspace{0.1em}]},  Cugny, B., Armandillo, E., and
  Karafolas, N., eds.,  {\bf 10564},  105642P, International Society for Optics
  and Photonics, SPIE (2017).

\end{thebibliography}
\bibliographystyle{spiebib} 

\end{document}